\documentclass[sigconf]{acmart}
\AtBeginDocument{%
  }

\copyrightyear{2026}
\acmYear{2026}
\setcopyright{cc}
\setcctype{by-nc-nd}
\acmDOI{}
\acmISBN{}

\acmBooktitle{}

\acmConference[CHI 2026 BiAlign Workshop]{CHI 2026 BiAlign Workshop}{April 15,  2026}{Barcelona, Spain}

\begin{document}

\title{Supporting Reflection in LLM-based Exploratory Search}

\author{Giulia Di Fede}
\orcid{0000-0003-3185-2727}
\affiliation{%
  \institution{Politecnico di Milano}
  \city{Milan}
  \country{Italy}
}
\email{giulia.difede@polimi.it}

\author{Salvatore Andolina}
\orcid{0000-0001-9804-4009}
\affiliation{%
  \institution{Politecnico di Milano}
  \city{Milan}
  \country{Italy}
}
\email{salvatore.andolina@polimi.it}

\begin{abstract}
Large Language Models (LLMs) can make exploratory search more efficient but may undermine the reflection and iterative sensemaking needed in unfamiliar domains. Existing LLM tools often prioritize rapid answers over supporting users in tracking how their understanding evolves and how well their strategies align with their goals. We present TrailLM, a system that helps users reconstruct and revisit their exploration paths to support reflection and metacognitive engagement during information seeking. By aligning LLM assistance with users' sensemaking workflows, TrailLM aims to preserve the benefits of LLM-based search while enhancing opportunities for critical reflection on one's own search process.
\end{abstract}

\ccsdesc[300]{Human-centered computing~Interaction design theory, concepts and paradigms}
\ccsdesc[500]{Human-centered computing~HCI theory, concepts and models}
\ccsdesc[300]{Information systems~Information retrieval}

\keywords{Exploratory Search, Critical Thinking, Meta-Cognition, Information Retrieval, Generative Artificial Intelligence, Large Language Models}

\maketitle
\section{Introduction}

Exploratory search is a cognitively demanding task characterized by vague and evolving information needs \cite{marchionini2006exploring}. Unlike lookup searches such as "How tall is Obama?", which rely on well-defined queries and can be answered instantly, exploratory search involves iterative query reformulation, multiple open tabs, and prolonged sessions that may span hours, days, or weeks. Tasks like planning a vacation or researching a new topic require users to discover new keywords, refine goals, and issue increasingly precise queries in a continuous, cognitively intense process of exploration and sensemaking \cite{Pirolli10032011}.

The advent of LLMs is transforming exploratory search, promising more efficient retrieval and summarization of relevant information. Yet users often struggle to translate their goals into the detailed prompts needed to use these systems effectively \cite{subramonyam2024bridging}. This is especially true in the early stages of exploratory search when intent is still vague and users are forming their needs and understanding of the information space. 
In more traditional exploratory search approaches, users iteratively refine queries, compare sources, and revise their information-seeking goals \cite{andolina2015intentstreams, andolina2018querytogether, klouche2015}. In LLM-based search, by contrast, users receive a direct answer to an initial, often underdeveloped prompt, which may encourage them to bypass the critical process of iterative sensemaking. Consequently, they may obtain incomplete or misleading responses that fail to match their actual needs \cite{alrabie2025towards}.
In addition, LLMs can generate false information that appears authoritative, known as hallucinations, and reinforce biases embedded in their training data. When users rely on LLMs without making the effort to iteratively make sense of the information space, they could face inhibition of reflection or critical thinking \cite{lee2025the, ford2026Rice} and incur the risk of more easily accepting information that is incomplete, inaccurate, or biased \cite{DiFede2026}.

To understand how to support reflection in LLM-based exploratory search, we focus on metacognition—the ability to reflect on and regulate one’s own thinking—which is increasingly central to designing \textit{aligned human–AI interactions} \cite{lee2025the, shen2024towards}. In exploratory search, users must reflect on their metacognitive processes to recognize what they know, identify knowledge gaps, and deliberately regulate their search. Effective exploratory search requires assessing the relevance of retrieved information, adapting queries, and refining mental models through ongoing reflection. Without such active engagement, users may passively accept information instead of critically evaluating and synthesizing it, becoming more vulnerable to misinformation and biased output.

We conducted a formative study to understand how people reflect on their own search strategies when exploring an unfamiliar topic through LLMs. 
Based on insights from the literature and the study, we designed \textbf{TrailLM}, a prototype intended to make LLM-based exploratory search more intentional through augmented interaction histories that help users reflect on their strategies and keep track of their current and future goals and subgoals for their search tasks.

\begin{figure*}[t!]
  \centering
  \includegraphics[width=0.80\linewidth]{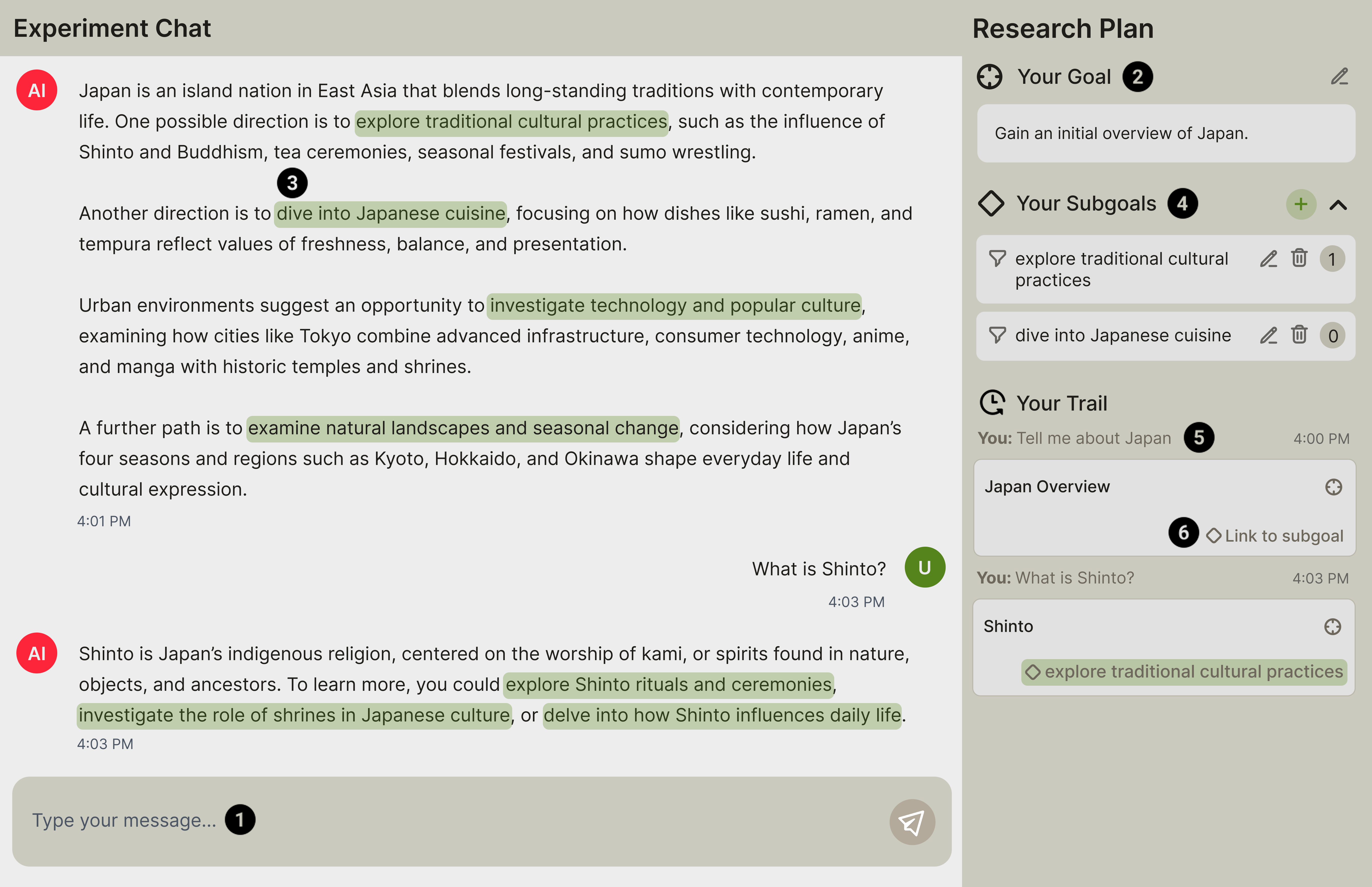}
  \Description{Screenshot of the TrailLM user interface showing input bar, exploration goal, LLM responses, subgoals section, and trail section.}
  \caption{The TrailLM user interface. 1) Input bar for submitting prompts; 2) exploration goal; 3) LLM responses with highlighted subgoal suggestions; 4) subgoals section for aggregating subgoals; 5) trail section showing interaction history; 6) linking trail tiles to subgoals and filtering by subgoal.
}~\label{fig:Interface}

\end{figure*}

\section{TrailLM}
\subsection{Design Goals}
To support meta-cognition and reflection in exploratory search tasks aided by LLMs, we designed TrailLM. In this section, we outline the design goals that motivated the system's key features.

\begin{itemize}
    \item \textit{Goal Alignment Monitoring}. Exploratory search is dynamic, with user goals evolving as new information appears. TrailLM shows how a user’s current path aligns with their self-defined subgoals by externalizing the system’s interpretation of those subgoals in a visible goal state. This makes active subgoals explicit, allowing users to reflect and adjust their exploration. Users can also revise or add subgoals at any time, so the system can better infer the current subgoal in later iterations and keep its assistance aligned.
    \item \textit{Discovering emerging exploration paths}. As exploratory search progresses, new directions and lines of inquiry can implicitly emerge through interaction. TrailLM supports users in identifying emerging paths by making alternative directions clearly visible.
    \item \textit{Facilitating information sense-making}. During exploratory search, users are exposed to vast and complex information spaces that imply non-linear information sense-making. By organizing the interaction history as a navigable trail, TraiLM allows users to revisit past steps of their exploration and connect them to different goals. 
\end{itemize}

In order to fulfill the above design goals, TraiLM was provided  with a user interface equipped with the affordances introduced in the next section.

\subsection{User Interface}
TrailLM's interface consists of a main chat area and a persistent right sidebar (Figure \ref{fig:Interface}). In the main chat, users interact with the LLM by entering prompts in the input bar at the bottom (Figure \ref{fig:Interface}-1). 
After a user submits a prompt, the system updates the exploration goal shown at the top of the sidebar (Figure \ref{fig:Interface}-2). This helps users reflect both on their current exploration path and on how well their prompts express their intended goals to the LLM. Users can revise the goal at any time using the modify button.
The LLM’s response in the main chat includes suggested subgoals that reflect possible directions for further exploration (Figure \ref{fig:Interface}-3). Clicking a suggestion adds it to the subgoals list in the sidebar (Figure \ref{fig:Interface}-4). Users can also manually add, edit, or delete subgoals to reorganize their exploration as it evolves.
Once a response is fully generated, a new trail tile appears in the sidebar’s trail section (Figure \ref{fig:Interface}-5). Selecting a tile returns the chat to that point, letting users revisit earlier steps. Users can tag each tile with one or more subgoals (Figure \ref{fig:Interface}-6), linking information to specific exploration paths. The trail can be filtered by subgoal so that only tiles associated with a chosen subgoal are shown, supporting both a global overview and focused views on particular directions.

\section{Conclusions}
We present TrailLM, a prototype designed to support reflection in LLM-based exploratory search. By making users aware of their search paths and pursued goals and subgoals, it helps them evaluate their progress and make more informed adjustments to their exploration plans. We are currently seeking feedback on the first prototype to identify potential improvements and ensure it meets user needs before moving forward. The next step will be to run a controlled user study. The findings from this study will allow us to confirm or revise our assumptions about how LLMs can best scaffold reflective, aligned interaction, and will guide future iterations of TrailLM and other alignment-centered interfaces.

\bibliographystyle{ACM-Reference-Format}
\bibliography{WORKING-BIB}
\end{document}